\documentclass{article}

\usepackage{arxiv}

\usepackage[utf8]{inputenc} %
\usepackage[T1]{fontenc}    %
\usepackage{hyperref}       %
\usepackage{url}            %
\usepackage{booktabs}       %
\usepackage{amsfonts}       %
\usepackage{nicefrac}       %
\usepackage{microtype}      %
\usepackage{graphicx}

\usepackage{multirow}%
\usepackage{amsmath,amssymb}%
\usepackage{mathtools}%
\usepackage{amsthm}%
\usepackage{mathrsfs}%
\usepackage[title]{appendix}%
\usepackage{xcolor}%
\usepackage{textcomp}%
\usepackage{manyfoot}%
\usepackage{algorithm}%
\usepackage{algorithmicx}%
\usepackage{algpseudocode}%
\usepackage{listings}%
\definecolor{codekeyword}{rgb}{0.0, 0.4, 0.65}
\definecolor{codecomment}{rgb}{0.4, 0.45, 0.5}
\definecolor{codestring}{rgb}{0.2, 0.5, 0.25}
\definecolor{codenumber}{rgb}{0.75, 0.75, 0.75}
\definecolor{codebase}{rgb}{0.15, 0.15, 0.15}
\usepackage{pgfplots}
\usepgfplotslibrary{fillbetween}
\usepgfplotslibrary{groupplots}
\pgfplotsset{compat=1.18}
\usepackage{comment}
\usetikzlibrary{spy}
\usepackage{xfrac}

\usepackage[version=4]{mhchem}
\usepackage{siunitx}
\usepackage{longtable,tabularx}

\newcolumntype{E}{S[table-format=1.3e-1]}
\newcolumntype{O}{S[table-format=-1.2]}
\newcolumntype{T}{S[table-format=1.3]}

\usepackage{tikz}
\pgfplotsset{compat=newest}

\definecolor{pv1}{rgb}{0.278431, 0.278431, 0.858824} %
\definecolor{pv2}{rgb}{0, 0, 0.360784}               %
\definecolor{pv3}{rgb}{0, 1, 1}                      %
\definecolor{pv4}{rgb}{0, 0.501961, 0}               %
\definecolor{pv5}{rgb}{1, 1, 0}                      %
\definecolor{pv6}{rgb}{1, 0.380392, 0}               %
\definecolor{pv7}{rgb}{0.419608, 0, 0}               %
\definecolor{pv8}{rgb}{0.878431, 0.301961, 0.301961} %

\pgfplotsset{
    colormap={papercmap}{
        rgb255(0cm)=(29, 100, 171)   %
        rgb255(1cm)=(255, 255, 255)  %
        rgb255(2cm)=(218, 66, 40)    %
    }
}

\theoremstyle{plain}
\theoremstyle{remark}
\theoremstyle{definition}

\title{Lattice Boltzmann Method for Compressible Navier--Stokes--Fourier Equations}

\author{
  Fedor Bukreev\thanks{These authors contributed equally to this work.} \\
  Lattice Boltzmann Research Group (LBRG) \\
  Institute of Mechanical Process Engineering and Mechanics (MVM) \\
  Karlsruhe Institute of Technology (KIT) \\
  Straße am Forum 8, 76131 Karlsruhe, Germany \\
  \texttt{fedor.bukreev@kit.edu} \\
  \And
  Adrian Kummerländer\footnotemark[1] \\
  Lattice Boltzmann Research Group (LBRG) \\
  Institute for Applied and Numerical Mathematics (IANM) \\
  Karlsruhe Institute of Technology (KIT) \\
  Englerstraße 2, 76131 Karlsruhe, Germany \\
  \texttt{kummerlaender@kit.edu} \\
  \And
  Mathias J. Krause\footnotemark[1] \\
  Lattice Boltzmann Research Group (LBRG) \\
  Institute of Mechanical Process Engineering and Mechanics (MVM) \\
  Institute for Applied and Numerical Mathematics (IANM) \\
  Karlsruhe Institute of Technology (KIT) \\
  Straße am Forum 8 / Englerstraße 2, 76131 Karlsruhe, Germany \\
  \texttt{mathias.krause@kit.edu} \\
}

\begin{document}
\maketitle

\begin{abstract}
A lattice Boltzmann scheme for the three-dimensional compressible Navier--Stokes--Fourier equations, derived automatically from the declared system by a symbolic compiler~\cite{pde2lbm}, is validated against exact solutions and published reference data.
The declared system carries the viscous stress and the heat flux as transported state, and is discretized on a D3Q7 lattice in single precision.
Against the exact Sod and Becker solutions the captured shock thickness converges at first order.
On the supersonic Taylor--Green vortex at $M_0 = 1.25$ the scheme at $512^3$ matches the reference dilatational dissipation more closely than seven compared solvers~\cite{Chapelier2024}, by thirty percent over the next best.
Every operator in this solver, the generated collision and constitutive closure and the added shock sensor alike, reads only the cell it acts on, and data reaches a neighbor only by streaming along the lattice characteristics.
\end{abstract}

\keywords{Compressible Navier--Stokes--Fourier \and LBM \and Turbulence \and Shock Capturing \and Taylor--Green Vortex}

\section{Introduction}
Compressible flow places two opposing demands on numerical models: that discontinuities are captured without oscillation and that a turbulent cascade is carried without being damped.
Dissipation settles both but with the caveat that each demand wants it in a place the other does not.
Finite-volume and finite-difference methods resolve this through approximate Riemann solvers, flux limiters and high-order reconstructions~\cite{Toro2009, Fu2016}, all of which gather extended neighborhoods.
The same gathering evaluates the velocity and temperature gradients that the viscous stress and the Fourier flux require.

The \emph{lattice Boltzmann method} (LBM)~\cite{Krueger2016, WolfGladrow1995} avoids it by construction, decomposing into a cell-local collision and a streaming step along discrete characteristics.
That locality is what carries the method onto heterogeneous supercomputers~\cite{Tolke2010, Godenschwager2013, Bauer2021walberla, Kummerlaender2023, Kummerlaender2026c}.
Standard scalar LBM does not extend to compressible flow without cost: matching the required moments calls for high-order Hermite equilibria~\cite{ShanYuanChen2006, Coreixas2017}, enlarged or multi-speed velocity sets~\cite{Frapolli2016}, an abandoned fixed lattice frame~\cite{Dorschner2018}, or a hybrid in which the energy equation is closed by a finite-difference solver alongside the lattice~\cite{Feng2019, Jacob2018}.
The last of these recovers compressibility at the price of the locality and efficiency that are one of the main motivations of the method.

A different approach assigns an independent set of distribution functions to each component of an extended macroscopic state and embeds the physical flux exactly in the first moment, so one compact lattice serves many systems~\cite{Bukreev2026, Guillon2024, Wissocq2025, pde2lbm}.
This is a discrete-kinetic relaxation approximation of conservation laws~\cite{JinXin1995, Natalini1998, AregbaNatalini2000, Bouchut1999}.
Automatic code generation is established for LBM, \emph{lbmpy} producing optimized kernels from a discrete-scheme specification~\cite{Hennig2023}.
The relaxation approximation is a single mechanical map, so it admits automation at the level of the declared system: a symbolic compiler derives the equilibrium, the auxiliary-variable cascade and the grid scaling from the conservation laws alone and emits hardware-saturating kernels for OpenLB~\cite{pde2lbm, Krause2021a}.
The compiler covers hyperbolic, parabolic and mixed systems, the compressible Navier--Stokes--Fourier equations among them, each verified by the \emph{method of manufactured solutions} (MMS)~\cite{pde2lbm}.

Manufactured fields are smooth by construction and driven by prescribed source.
Verification against them establishes that the generated scheme converges to the system it was asked to discretize but not whether a scheme captures shocks or transports a supersonic cascade.
Testing this requires comparison to physical benchmarks.
Accordingly, the relaxed-flux form adopted here was derived by hand in~\cite{Bukreev2026} and evaluated there on magnetohydrodynamic benchmarks including the Brio--Wu shock tube and the Orszag--Tang vortex.

Following up on~\cite{pde2lbm}, the present work supplies that validation for the compressible Navier--Stokes--Fourier system.
It is declared to the compiler in that form, carrying the viscous stress and the heat flux as transported state.
The compiler presently carries no shock-capturing mechanism and controls dissipation only through the scalar relaxation time, which is why we add shock capturing \emph{by hand} using OpenLB's established operator concept~\cite{Kummerlaender2026c}.
Commonly sensors rest on pressure-based switches~\cite{Jameson1981}, on gates comparing dilatation against vorticity magnitude~\cite{Ducros1999}, or on localised artificial diffusivity driven by a gradient indicator~\cite{CookCabot2005, KawaiLele2008}.
The sensor used here recovers the local dilatation from the cell's own moments and reads only the cell it acts on.

The benchmarks are chosen to separate the two extreme modes.
The Sod problem~\cite{Sod1978} validates captured discontinuities against an exact Riemann solution in the near-inviscid limit.
Becker's traveling wave~\cite{Becker1922, MorduchowLibby1949} has a smooth closed-form profile and tests the coupled stress and heat-flux transport under a strong thermodynamic gradient.
The compressible Taylor--Green vortex~\cite{TaylorGreen1937, Brachet1983} at $\text{Re} = 1600$~\cite{Wang2013, DeBonis2013} and $M_0 = 1.25$~\cite{PengYang2018, Chapelier2024} combines both, generating shocklets inside a decaying cascade, and is validated against the seven-solver comparison and $2048^3$ reference of Chapelier et al.~\cite{Chapelier2024}.

Lattice Boltzmann results on the incompressible form of that vortex are numerous~\cite{Kajzer2014, Ning2016, Nathen2018, Haussmann2019, Negro2019, ChavezModena2020, Geier2021, Simonis2021, Mimeau2021, Simonis2022, Hennig2023, Strzelczyk2024}.
Compressible results cover hybrid formulations~\cite{Zhao2020, Boivin2021, Vienne2024}, off-lattice semi-Lagrangian schemes up to $M = 2$~\cite{Wilde2020, Wilde2021a, Wilde2021b}, and discrete-velocity and gas-kinetic schemes at $M = 1.25$~\cite{Bo2017, Cao2022, Guo2023}.
Shock capturing on the case has been assessed for finite-difference schemes~\cite{LusherSandham2021}.

The remainder of this paper is organized as follows.
Section~\ref{sec:methodology} states the declared system, its discretisation on the D3Q7 lattice and the shock sensor.
Section~\ref{sec:validation} reports the two shock tubes against their exact solutions in Section~\ref{sec:shocktubes}, and the compressible Taylor--Green vortex against the multi-code comparison in Section~\ref{sec:ctgv}.
\section{Methodology}\label{sec:methodology}

The derivation of the scheme, its Chapman--Enskog consistency, the sub-characteristic admissibility condition, the automated grid scaling and the reference-shifted formulation used in single precision are given in~\cite{pde2lbm} and are not repeated here.
This section states what is specific to the present target, the system declared and the shock capturing that the physical benchmarks require and the compiler does not supply.

\subsection{Target System}\label{sec:declared}

The compiler takes a system of conservation laws as its sole entry point~\cite{pde2lbm},
\begin{equation}
\frac{\partial \mathbf{Q}}{\partial t} + \nabla \cdot \mathbf{\Phi} = \mathbf{S}
\end{equation}
with $\mathbf{Q}$ the macroscopic state vector, $\mathbf{\Phi}$ the physical flux tensor and $\mathbf{S}$ a local source.
The compressible viscous gas is declared here in the relaxed-flux form of~\cite{Bukreev2026}, which carries the viscous stress $\boldsymbol{\tau}$ and the heat flux $\mathbf{q}$ as transported state rather than assembling them from tracked gradients of the conserved fields, giving fourteen components in three dimensions:
\begin{align}\label{eq:system}
    \mathbf{Q} = \begin{pmatrix*}[l]
    \rho \\
    \rho \mathbf{u} \\
    E \\
    \boldsymbol{\tau} \\
    \mathbf{q}
    \end{pmatrix*}, \quad
    \mathbf{\Phi} = \begin{pmatrix*}[l]
    \rho \mathbf{u} \\
    \rho \mathbf{u} \otimes \mathbf{u} + p \mathbf{I} - \boldsymbol{\tau} \\
    (E + p)\mathbf{u} - \boldsymbol{\tau} \cdot \mathbf{u} + \mathbf{q} \\
    \boldsymbol{\tau} \otimes \mathbf{u} - \frac{\mu}{\tau_R}\,\mathbf{\Lambda}(\mathbf{u}) \\
    \mathbf{q} \otimes \mathbf{u} + \frac{\kappa}{\tau_q}\,T\,\mathbf{I}
    \end{pmatrix*}, \quad
    \mathbf{S} = \begin{pmatrix*}[l]
    0 \\
    \mathbf{0} \\
    0 \\
    -\boldsymbol{\tau}/\tau_R \\
    -\mathbf{q}/\tau_q
    \end{pmatrix*}.
\end{align}
Here $\rho$ is the density, $\mathbf{u}$ the velocity, $E$ the total energy, $\boldsymbol{\tau}$ the symmetric viscous stress with six independent components and $\mathbf{q}$ the heat flux.
The pressure follows the ideal gas law $p = (\gamma - 1)(E - \tfrac{1}{2}\rho|\mathbf{u}|^2)$ with adiabatic index $\gamma$ and temperature $T = p/\rho$, and $\mathbf{\Lambda}(\mathbf{u})$ collects the velocity contributions whose divergence reproduces the deviatoric strain rate $\nabla\mathbf{u} + (\nabla\mathbf{u})^T - \tfrac{2}{3}(\nabla\cdot\mathbf{u})\mathbf{I}$.

Listing~\ref{lst:declaration} gives this system as it is declared to the compiler.

\begin{lstlisting}[language=Python, float=htbp, label={lst:declaration},
caption={The system of~\eqref{eq:system} as declared to the compiler.
The stress is one symmetric tensor state and the heat flux one vector state, so the fourteen components are declared as five equations.
The equilibrium, the auxiliary-variable cascade, the grid scaling and the collide-and-stream kernel follow from this declaration alone.}]
from pde2lbm import *

eqs = ConservationLaws(dim=3, lattice="D3Q7")

# 1. Conserved state, with the viscous stress and the heat flux carried alongside it
rho  = eqs.state("rho",  dim=mass / length**3)
rhou = eqs.state("rhou", shape=3, dim=momentum / length**3)
E    = eqs.state("E",    dim=energy / length**3)
tau  = eqs.state("tau",  shape=(3, 3), symmetric=True, dim=pressure)
q    = eqs.state("q",    shape=3, dim=energy / length**2 / time)

# 2. Parameters
gamma = eqs.parameter("gamma")
mu    = eqs.parameter("mu",    dim=pressure * time)
kappa = eqs.parameter("kappa", dim=pressure * time)
tau_R = eqs.parameter("tau_R", dim=time)
tau_q = eqs.parameter("tau_q", dim=time)
T_ref = eqs.parameter("T_ref", dim=energy / mass)

# 3. Dependent variables
u    = rhou / rho
p    = (gamma - 1) * (E - 0.5 * rho * u.dot(u))
T    = p / rho
T_S  = 0.4042                                             # Sutherland temperature, in units of T_ref
mu_T = mu * (1 + T_S) * (T / T_ref)**Rational(3, 2) / (T / T_ref + T_S)
visc = mu_T / tau_R
heat = kappa * (mu_T / mu) / tau_q

# 4. Equation system
eqs.add([
    Eq(dt(rho)  + div(rhou)                                  , 0),
    Eq(dt(rhou) + div(outer(rhou, u) + p * eye(3) + tau)      , zeros(3, 1)),
    Eq(dt(E)    + div((E + p) * u + tau * u + q)              , 0),
    Eq(dt(tau)  + div(outer(tau, u) + visc * dev_strain(u))   , -tau / tau_R),
    Eq(dt(q)    + div(outer(q, u) + heat * T * eye(3))        , -q / tau_q),
])

# 5. Generate the OpenLB Dynamics
eqs.compile(class_name="CompressibleNSF")
\end{lstlisting}

The stress and heat-flux components are therefore not diagnostic but transported and relaxed toward the law that would otherwise define them, a relaxation approximation in the sense of~\cite{JinXin1995, Natalini1998} applied at the level of the constitutive fluxes.
Multiplying their rows by the corresponding relaxation time exhibits the structure directly, $\tau_R\left(\partial_t\boldsymbol{\tau} + \nabla\cdot(\boldsymbol{\tau}\otimes\mathbf{u})\right) + \boldsymbol{\tau} = \mu\,\nabla\cdot\mathbf{\Lambda}$ and $\tau_q\left(\partial_t\mathbf{q} + \nabla\cdot(\mathbf{q}\otimes\mathbf{u})\right) + \mathbf{q} = -\kappa\nabla T$, whose $\tau_R, \tau_q \to 0$ limit is the Newtonian and Fourier closure.
Both relaxation times are free parameters rather than modelled quantities, since the recovered viscosity and conductivity are $\mu$ and $\kappa$ independently of them.
They set only the size of the $\mathcal{O}(\tau)$ deviation from the Navier--Stokes--Fourier system.
On the shock tubes they are tied to the timestep and that deviation vanishes under refinement, whereas on the Taylor--Green vortex they are held at a fixed physical value chosen for accuracy.

\subsection{Discretisation}\label{sec:discretisation}

The declared system is discretized by the generated lattice Boltzmann integration of~\cite{pde2lbm}, which allocates an independent set of populations to each of the fourteen components and embeds the flux $\mathbf{\Phi}_k$ exactly in the first moment of a linear equilibrium.
It recovers $Q_k$ from the zeroth moment alone and marches the result by a BGK collide-and-stream update with the source projected onto the lattice weights.
The present target uses the D3Q7 stencil, for which $c_s^2 = 1/4$.
The throughput of this kernel class is characterized in~\cite{pde2lbm} for the same fourteen-field system.
Every result reported here was computed with single-precision populations in the reference-shifted formulation of~\cite{pde2lbm}.

\subsection{Local Shock Capturing}\label{sec:sensor}

A lattice scheme that aims to stay strictly local during collision cannot capture shocks with the usual Jameson-type sensor~\cite{Jameson1981}, which needs second differences of the pressure and therefore neighbor access.
The sensor described here reads only the cell it applies to.

The extended state carries the viscous stresses and heat fluxes, so each field group has its own set of populations and its own non-equilibrium part.
The local dilatation $\nabla\cdot\mathbf{u}$ is therefore already present in the data a cell holds, and is recovered from it by an algebraic expression in that cell's own moments, without differencing the grid.
The sensor requires only that the recovery be local and free of relaxation lag.

Shocks compress, so the sensor is one-sided.
The compression is normalized by the lattice sound speed, which turns it into a per-cell compression Mach number
\begin{equation}
    s = \frac{\max(0, -\nabla\cdot\mathbf{u})}{c_s},
\end{equation}
and above a threshold $J_{\min}$ on that normalized quantity the sensor returns
\begin{equation}
    \chi = \begin{cases}
        \mathrm{clamp}\!\left( C_{\!\sigma}\, s, 0, 1 \right) & s > J_{\min},\\
        0 & \text{otherwise},
    \end{cases}
\end{equation}
and $\chi$ then lowers the relaxation rate of that cell from its baseline $\omega$ toward a floor $\omega_{\min}$,
\begin{equation}
    \omega_h = \omega - \chi\,\left(\omega - \omega_{\min}\right).
\end{equation}
That baseline $\omega = 1/\tau_{\rm base}$ is itself set by a background dissipation factor $f_\mu$,
\begin{equation}\label{eq:fmu}
    \tau_{\rm base} = \frac{1}{2} + 4 f_\mu \frac{\mu}{\rho}\frac{\Delta t}{\Delta x^2},
\end{equation}
which on the D3Q7 lattice adds a numerical viscosity $\nu_{\rm num} = f_\mu\,\mu/\rho$ exactly, so $f_\mu$ is the mesh-independent ratio of numerical to molecular viscosity.
The physical viscosity is carried by the stress flux rather than by this rate, so $f_\mu$ is a purely numerical control and $f_\mu = 0$ is admissible wherever the scheme is stable without it.

The blend above is part of the generated collision: the specification declares $\chi$ as an input field and the emitted kernel applies it to the relaxation rates.
The evaluation of $\chi$ itself is not generated and is supplied as a companion operator.

Table~\ref{tab:parameters} collects the parameters of the scheme.

\begin{table}[htbp]
    \centering
    \caption{Parameters of the scheme, grouped by the subsection that introduces them.}
    \label{tab:parameters}
    \begin{tabularx}{\linewidth}{@{}l>{\raggedright\arraybackslash}X@{}}
        \toprule
        \multicolumn{2}{@{}l}{\textit{Declared system} (Section~\ref{sec:declared})} \\
        \cmidrule(r{\tabcolsep}){1-1}\cmidrule(l{\tabcolsep}){2-2}
        $\tau_R$ & Relaxation time of the viscous stress $\boldsymbol{\tau}$ toward the Newtonian closure. \\
        $\tau_q$ & Relaxation time of the heat flux $\mathbf{q}$ toward the Fourier closure. \\
        \midrule
        \multicolumn{2}{@{}l}{\textit{Discretisation} (Section~\ref{sec:discretisation})} \\
        \cmidrule(r{\tabcolsep}){1-1}\cmidrule(l{\tabcolsep}){2-2}
        $\mathrm{CFL}$ & Acoustic Courant number, which fixes $\Delta t$ at a given $\Delta x$. \\
        \midrule
        \multicolumn{2}{@{}l}{\textit{Local shock capturing} (Section~\ref{sec:sensor})} \\
        \cmidrule(r{\tabcolsep}){1-1}\cmidrule(l{\tabcolsep}){2-2}
        $f_\mu$ & Numerical viscosity added by the baseline rate~\eqref{eq:fmu}, in multiples of the molecular value. \\
        $C_{\!\sigma}$ & Slope of the sensor response, saturating at a compression Mach number $s = 1/C_{\!\sigma}$. \\
        $J_{\min}$ & Threshold on the compression Mach number, below which the sensor is inactive. \\
        $\omega_{\min}$ & Floor on the relaxation rate, which bounds the damping the sensor can apply. \\
        \bottomrule
    \end{tabularx}
\end{table}

\section{Validation}\label{sec:validation}

Every result in this paper is generated from one specification.
The Taylor--Green vortex uses Sutherland's law, while both shock tubes use a constant $\mu$, because the Becker profile they are scored against is itself a constant-$\mu$ solution (Section~\ref{sec:shocktubes}).

\subsection{Shock Tubes}\label{sec:shocktubes}

Both shock tubes are evaluated against their exact solutions, the Riemann solution for Sod and the Becker traveling wave for the viscous case.

\paragraph{Viscosity of the Becker benchmark.}
The Becker profile at $\mathrm{Pr} = 3/4$ serves as both the initial condition and the reference.
Its closed form is the exact integral of the shock-frame momentum balance, and that integral is elementary only for a constant dynamic viscosity.
The profile is therefore a constant-$\mu$ solution and the simulations use a constant $\mu$.
Matching it also fixes the magnitude through $\mathrm{Re}_s = \rho_0 c_0 M_s L/\mu$, so at $M_s = 2$ and $\mathrm{Re}_s = 1000$ on unit scales the value is $\mu = 2\times 10^{-3}$.

\paragraph{Configuration.}
Both tubes use $\gamma = 1.4$ and the sensor of Section~\ref{sec:sensor} at $C_{\!\sigma} = 20$, $J_{\min} = 3\cdot10^{-3}$ and $\omega_{\min} = 1$.
The inviscid case runs at $\mu = 10^{-6}$ and $\mathrm{CFL} = 0.35$, the viscous case at $\mathrm{Pr} = 3/4$ and $\mathrm{CFL} = 0.125$.
The relaxation times are tied to the timestep, $\tau_R = \tau_q = 5\,\Delta t$, so the $\mathcal{O}(\tau)$ modeling error falls with the mesh.

\begin{figure}[htbp]
    \centering
    \includegraphics{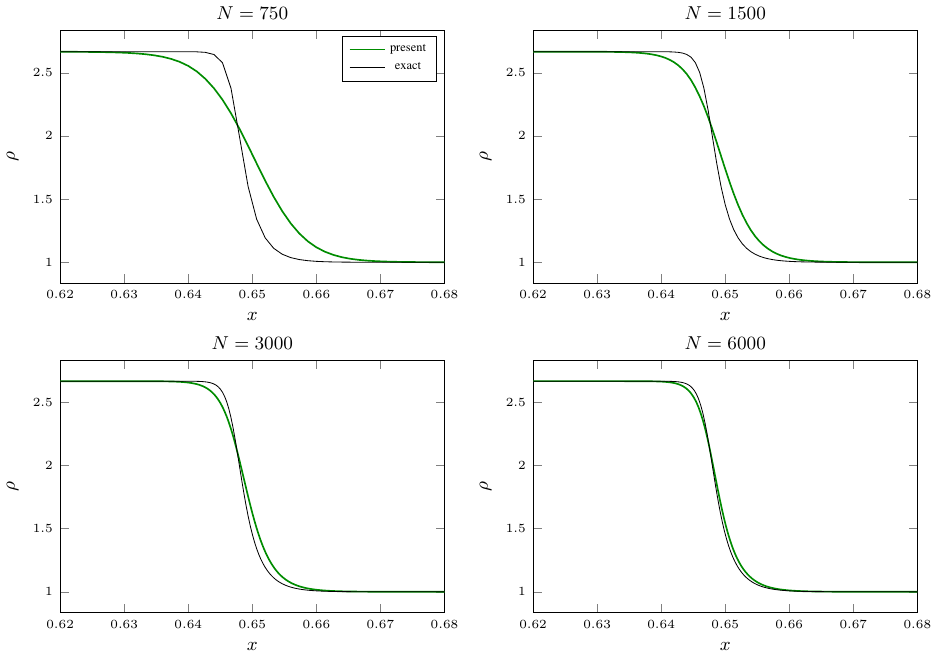}
    \caption{Viscous shock tube, density profile against the exact Becker traveling wave at $t=0.2$.}
    \label{fig:becker_profiles}
\end{figure}

\begin{figure}[htbp]
    \centering
    \includegraphics{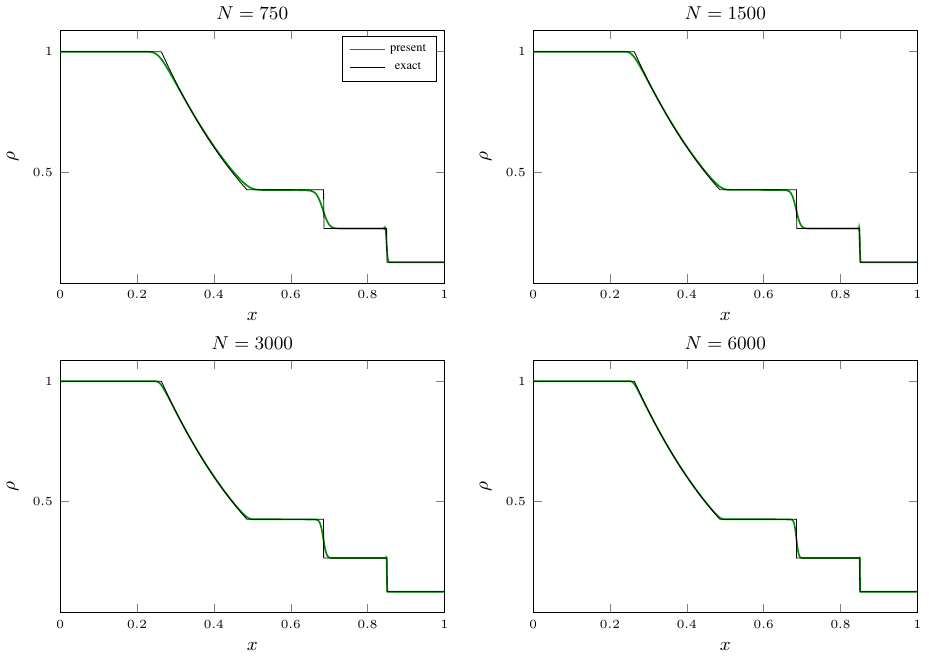}
    \caption{Sod shock tube, density profile against the exact Riemann solution at $t=0.2$.}
    \label{fig:sod_profiles}
\end{figure}

\paragraph{Norms.}
The exact solution of the inviscid case is discontinuous, so with a shock captured in a fixed number of cells the attainable orders are first in $L_1$, one half in $L_2$ and zero in $L_\infty$.
$L_1$ is the appropriate measure and all three are tabulated in Tables~\ref{tab:becker_conv} and~\ref{tab:sod_conv}, with the density profiles in Figures~\ref{fig:becker_profiles} and~\ref{fig:sod_profiles}.
The viscous case has a smooth exact solution, carries no such ceiling, and there all three norms agree.
The captured shock thickness converges at first order (Table~\ref{tab:thickness}) and the norms converge more slowly still.

\begin{table}[htbp]
    \centering
    \caption{Captured shock thickness, measured on the density profile between the two levels bracketing the transition.
    The Sod shock is a true discontinuity, so its thickness is pure numerical smearing and holds a fixed cell count, that is $w \propto \Delta x$.
    The Becker shock has a physical thickness $w_{\rm ex} = 7.554\cdot10^{-3}$, and the excess $|w/w_{\rm ex} - 1|$ falls at first order.}
    \label{tab:thickness}
    \begin{tabular}{r E O E T O}
\toprule
& \multicolumn{2}{c}{Sod} & \multicolumn{3}{c}{Becker} \\
\cmidrule(lr){2-3} \cmidrule(lr){4-6}
$N$ & {$w$} & {$w/\Delta x$} & {$w$} & {$w/w_{\rm ex}-1$} & {order} \\
\midrule
    750 & 6.345e-03 & 4.77 & 1.914e-02 & 1.534 & {--} \\
    1500 & 3.162e-03 & 4.75 & 1.309e-02 & 0.732 & 1.07 \\
    3000 & 1.639e-03 & 4.92 & 1.019e-02 & 0.348 & 1.07 \\
    6000 & 8.048e-04 & 4.83 & 8.817e-03 & 0.167 & 1.06 \\
\bottomrule
\end{tabular}

\end{table}

\begin{table}[htbp]
    \centering
    \caption{Viscous shock tube, mesh convergence against the exact Becker traveling wave.
    The exact solution is smooth, so all three norms converge.}
    \label{tab:becker_conv}
    \small
\setlength{\tabcolsep}{3pt}
\begin{tabular}{r EO EO EO EO EO}
\toprule
$N$ & \multicolumn{2}{c}{$L_1(\rho)$} & \multicolumn{2}{c}{$L_2(\rho)$} & \multicolumn{2}{c}{$L_\infty(\rho)$} & \multicolumn{2}{c}{$L_1(p)$} & \multicolumn{2}{c}{$L_1(u)$} \\
\cmidrule(lr){2-3} \cmidrule(lr){4-5} \cmidrule(lr){6-7} \cmidrule(lr){8-9} \cmidrule(lr){10-11}
\midrule
    750 & 4.130e-03 & {--} & 1.871e-02 & {--} & 1.697e-01 & {--} & 4.320e-03 & {--} & 6.533e-03 & {--} \\
    1500 & 2.024e-03 & 1.03 & 1.051e-02 & 0.83 & 1.102e-01 & 0.62 & 2.122e-03 & 1.03 & 3.149e-03 & 1.05 \\
    3000 & 1.012e-03 & 1.00 & 5.569e-03 & 0.92 & 6.303e-02 & 0.81 & 1.038e-03 & 1.03 & 1.519e-03 & 1.05 \\
    6000 & 6.071e-04 & 0.74 & 2.846e-03 & 0.97 & 3.300e-02 & 0.93 & 5.609e-04 & 0.89 & 7.784e-04 & 0.96 \\
\bottomrule
\end{tabular}

\end{table}

\begin{table}[htbp]
    \centering
    \caption{Sod shock tube, mesh convergence against the exact Riemann solution.}
    \label{tab:sod_conv}
    \small
\setlength{\tabcolsep}{3pt}
\begin{tabular}{r EO EO EO EO EO}
\toprule
$N$ & \multicolumn{2}{c}{$L_1(\rho)$} & \multicolumn{2}{c}{$L_2(\rho)$} & \multicolumn{2}{c}{$L_\infty(\rho)$} & \multicolumn{2}{c}{$L_1(p)$} & \multicolumn{2}{c}{$L_1(u)$} \\
\cmidrule(lr){2-3} \cmidrule(lr){4-5} \cmidrule(lr){6-7} \cmidrule(lr){8-9} \cmidrule(lr){10-11}
\midrule
    750 & 7.941e-03 & {--} & 1.764e-02 & {--} & 9.119e-02 & {--} & 5.882e-03 & {--} & 1.173e-02 & {--} \\
    1500 & 5.009e-03 & 0.66 & 1.373e-02 & 0.36 & 8.815e-02 & 0.05 & 3.389e-03 & 0.80 & 6.583e-03 & 0.83 \\
    3000 & 3.193e-03 & 0.65 & 1.109e-02 & 0.31 & 8.706e-02 & 0.02 & 1.955e-03 & 0.79 & 3.742e-03 & 0.81 \\
    6000 & 2.042e-03 & 0.64 & 8.954e-03 & 0.31 & 8.367e-02 & 0.06 & 1.094e-03 & 0.84 & 2.052e-03 & 0.87 \\
\bottomrule
\end{tabular}

\end{table}

\subsection{Compressible Taylor--Green Vortex}\label{sec:ctgv}

The compressible Taylor--Green vortex couples the thermodynamic and velocity fields strongly enough that dilatation becomes a leading-order effect.
As the vortical structures break down, localised compression generates shocklets inside an otherwise smooth cascade, and the sensor of Section~\ref{sec:sensor} must act on the shocklets there without removing energy from the solenoidal motion around them.

The case is initialized at a nominal Mach number $M = 1.25$, a Reynolds number $\text{Re} = 1600$ and a Prandtl number $\text{Pr} = 0.71$, with $\gamma = 1.4$, $\rho_\infty = u_0 = L = 1$, $\tau_R = \tau_q = 5\cdot10^{-3}$ and $\mathrm{CFL} = 0.1$.
Following Chapelier et al.~\cite{Chapelier2024} the initial temperature is a uniform reference $T_0$, so the initial density follows the pressure as $\rho = p/T_0$, and the velocity and pressure fields retain the standard periodic Taylor--Green form without artificial noise.
The flow is characterized over the convective time $t_c = t\,u_0/L$ up to $t_c \le 20$ by the kinetic energy $E_k$ and by the split of the dissipation rate into its solenoidal and dilatational parts,
\begin{equation}
    \varepsilon_s = \frac{1}{\rho_{\text{ref}}} \left\langle \mu_{\text{local}} \, \|\nabla \times \mathbf{u}\|^2 \right\rangle,
    \qquad
    \varepsilon_d = \frac{4}{3} \frac{1}{\rho_{\text{ref}}} \left\langle \mu_{\text{local}} \, (\nabla \cdot \mathbf{u})^2 \right\rangle,
\end{equation}
where $\langle \cdot \rangle$ is the volume average over the periodic box and $\mu_{\text{local}}$ follows the Sutherland law of the benchmark~\cite{Chapelier2024}, $\mu(T) = 1.4042\,(T/T_0)^{3/2} / (T/T_0 + 0.4042)\,\mu$.
The two measures separate the vortical cascade from the compression the shocklets carry, and the sensor has to respect that separation.
Because the mesh is uniform and the domain periodic, both curl and divergence are evaluated spectrally rather than by finite differences, and $\mu_{\text{local}}$ is applied per cell in physical space.

The scheme and its sensor are those of Section~\ref{sec:sensor}.
This section examines how the sensor behaves on a flow that contains both shocklets and a cascade.

\begin{figure}[htbp]
    \centering
    \includegraphics{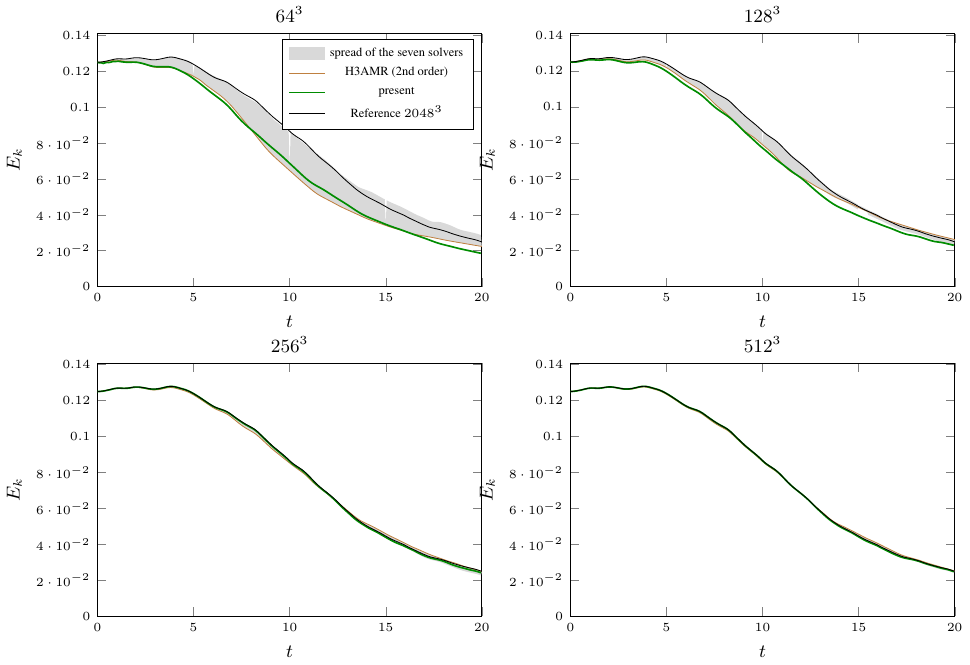}
    \caption{Time evolution of kinetic energy.
    The gray band spans the seven reference solvers, with H3AMR shown separately as the only second-order scheme among them.
    Reference: $2048^3$ sixth-order TENO.}
    \label{fig:ctgv_ek}
\end{figure}

\begin{figure}[htbp]
    \centering
    \includegraphics{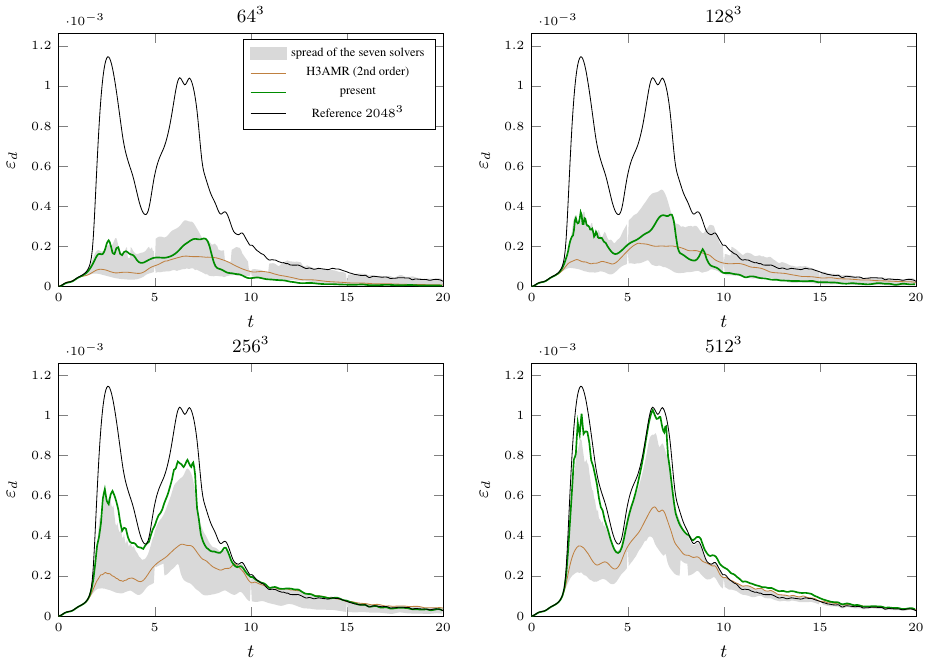}
    \caption{Time evolution of dilatational dissipation.
    The gray band spans the seven reference solvers, with H3AMR shown separately as the only second-order scheme among them.
    Reference: $2048^3$ sixth-order TENO.}
    \label{fig:ctgv_epsd}
\end{figure}

\begin{figure}[htbp]
    \centering
    \includegraphics{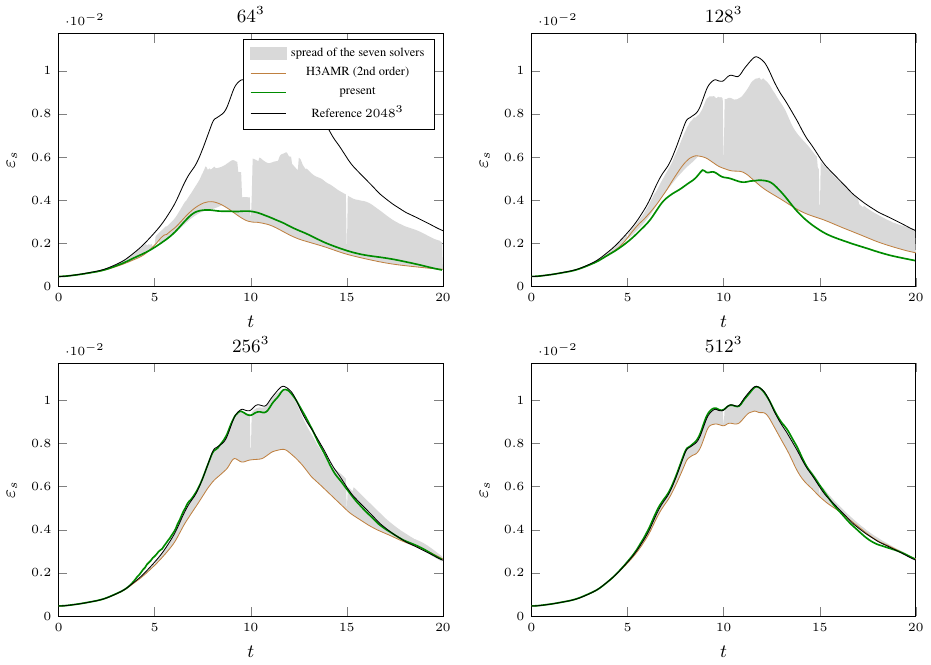}
    \caption{Time evolution of solenoidal dissipation.
    The gray band spans the seven reference solvers, with H3AMR shown separately as the only second-order scheme among them.
    Reference: $2048^3$ sixth-order TENO.}
    \label{fig:ctgv_epss}
\end{figure}

\begin{figure}[htbp]
    \centering
    \includegraphics{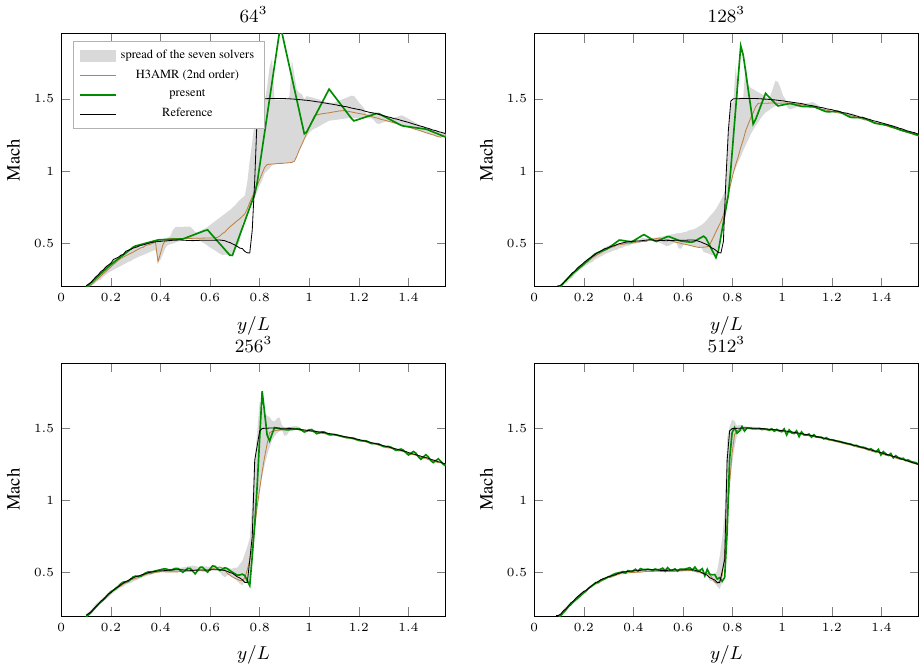}
    \caption{Mach profiles along the $y$ line at $x=z=0$ at $t_c=2.5$, the instant of peak dilatational dissipation.
    $L$ is the vortex reference length, so the periodic box has side $2\pi L$ and the panels show the first quarter of the $y$ line.}
    \label{fig:ctgv_mach}
\end{figure}

Three parameters are held fixed across the comparison of Figures~\ref{fig:ctgv_ek}--\ref{fig:ctgv_mach} and Table~\ref{tab:ctgv_l2}, namely $C_{\!\sigma} = 40$, $J_{\min} = 10^{-2}$ and $\omega_{\min} = 1.987$.
The grid-convergence study of Table~\ref{tab:ctgv_conv} is a separate configuration at the lower $C_{\!\sigma} = 20$, and is labeled as such.
The background dissipation is $f_\mu = 10^{-2}$ at $256^3$ and $512^3$, and $f_\mu = 1.8$ and $0.9$ on the $64^3$ and $128^3$ panels.

\begin{figure}[htbp]
    \centering
    \includegraphics[width=\textwidth]{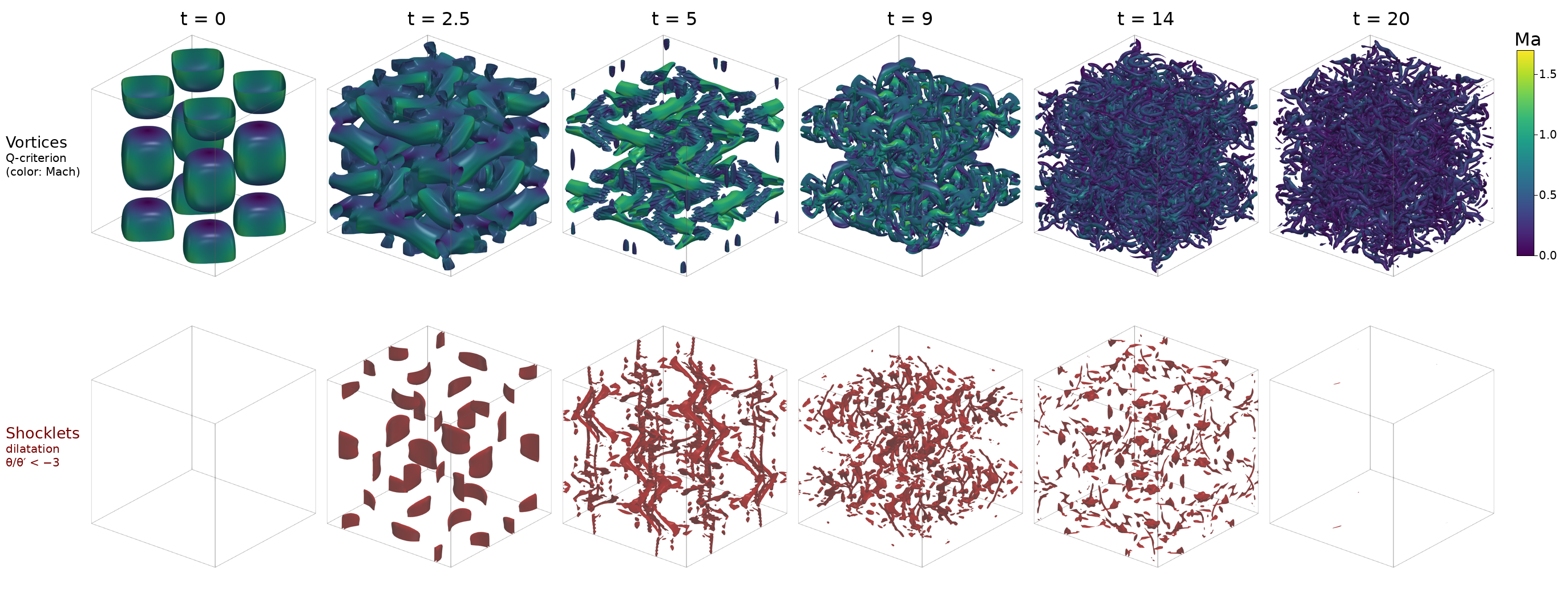}
    \caption{Flow visualization of the present $256^3$ result ($\text{Re} = 1600$, $M_0 = 1.25$) at $t_c \in \{0, 2.5, 5, 9, 14, 20\}$.
    Top: vortices as $Q$-criterion iso-surfaces colored by the local Mach number.
    Bottom: shocklets as iso-surfaces of the dilatation $\theta = \nabla\cdot\mathbf{u}$ with $\theta/\theta' < -3$~\cite{PengYang2018}.}
    \label{fig:ctgv_viz}
\end{figure}

Figure~\ref{fig:ctgv_viz} shows the two features the sensor must reconcile on this flow, the vortical cascade and the shocklets that form on the symmetry planes.
Vortices are rendered as iso-surfaces of the $Q$-criterion colored by the local Mach number, and shocklets as iso-surfaces of the dilatation $\theta = \nabla\cdot\mathbf{u}$ satisfying $\theta/\theta' < -3$~\cite{PengYang2018}, with $\theta'$ the r.m.s.\ dilatation.
The initial field is solenoidal, so no shocklets are present at $t_c = 0$.
They form through the transition, are most numerous near $t_c = 5$ to $9$, and decay by $t_c = 20$.
The local Mach number exceeds the free-stream value of $1.25$ in a small fraction of the volume, with under one percent of it above $M = 1.5$, in the expansion regions that seed the shocklets.

\paragraph{Comparison with the reference solvers.}

Figures~\ref{fig:ctgv_ek}--\ref{fig:ctgv_mach} compare the scheme against the seven solvers collected by Chapelier et al.\ and against their $2048^3$ TENO reference, with the gray band spanning the seven.
The seven are CODA~\cite{CODA}, NS3D~\cite{NS3D}, FLEXI~\cite{FLEXI}, OpenSBLI~\cite{Lusher2021}, SD3D~\cite{SD3D}, SPADE~\cite{SPADE} and H3AMR~\cite{H3AMR}.
H3AMR is drawn separately as the only second-order scheme among them, and since the generated schemes of~\cite{pde2lbm} are verified at second order by the method of manufactured solutions it is the like-for-like comparison here.
Every error quoted in this subsection and the next two is a relative $L_2$ over $t_c \in [0,20]$ against that reference, with the present result and the seven solvers scored by the same routine.
The reference and solver curves are extracted from the figures of that work.
Chapelier et al.\ note that the first dilatational peak is not mesh converged even at $2048^3$, since the shocks continue to sharpen, and for that peak the reference is itself resolution dependent.
The subsonic case below runs only to $t_c = 10$ and is scored over that interval.

\begin{table}[htbp]
    \centering
    \caption{Relative $L_2$ error against the $2048^3$ reference over $t_c \in [0,20]$ at $N=256$.}
    \label{tab:ctgv_l2}
    \begin{tabular}{lccc}
\toprule
& $E_k$ & $\varepsilon_s$ & $\varepsilon_d$ \\
\midrule
    \textbf{Present} & 0.0053 & 0.0224 & 0.3370 \\
    CODA & 0.0018 & 0.0239 & 0.6009 \\
    NS3D & 0.0029 & 0.0284 & 0.3721 \\
    FLEXI & 0.0022 & 0.0279 & 0.5095 \\
    OpenSBLI & 0.0035 & 0.0448 & 0.5623 \\
    SD3D & 0.0136 & 0.0896 & 0.8047 \\
    SPADE & 0.0076 & 0.0901 & 0.6918 \\
    H3AMR & 0.0119 & 0.2205 & 0.6721 \\
\bottomrule
\end{tabular}

\end{table}

\begin{figure}[htbp]
    \centering
    \includegraphics{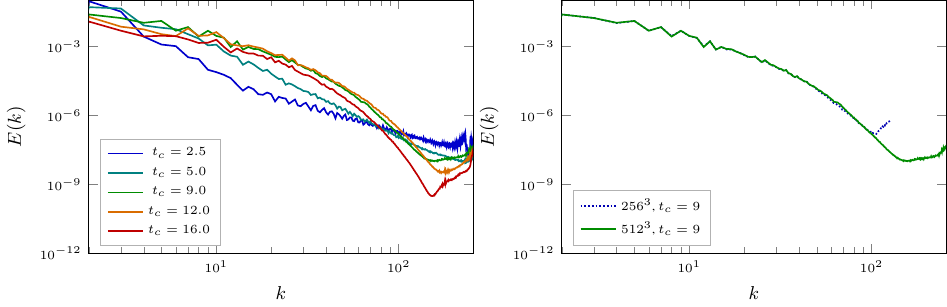}
    \caption{Shell-averaged velocity spectra at $512^3$, normalized so that $\sum_k E(k) = \langle |u|^2 \rangle / 2$.
    Left: through the run.
    Right: resolution comparison at $t_c=9$, near the peak of the solenoidal dissipation.}
    \label{fig:ctgv_spectrum}
\end{figure}

\begin{figure}[htbp]
    \centering
    \includegraphics{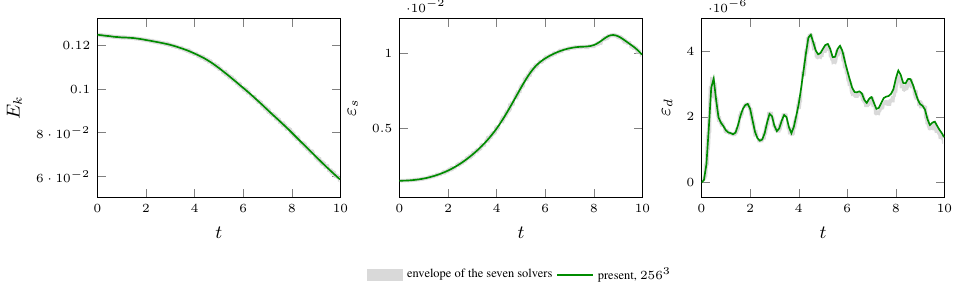}
    \caption{Subsonic verification case, $\text{Re}=500$, $M_0=0.5$.
    The spread of all seven reference solvers is shown as an envelope.}
    \label{fig:ctgv_re500}
\end{figure}

\paragraph{Sensor slope.}

$C_{\!\sigma}$ trades the two dissipation errors against one another, swept here at $512^3$ from $10$ to $80$.

The solenoidal error saturates at $C_{\!\sigma}=40$ and the dilatational error is minimized at $C_{\!\sigma}=20$ (Table~\ref{tab:ctgv_gain}).
The figures show $C_{\!\sigma}=40$.

\begin{table}[htbp]
    \centering
    \caption{Sensor slope swept at $512^3$.
    Relative $L_2$ against the $2048^3$ reference over $t_c \in [0,20]$.}
    \label{tab:ctgv_gain}
    \begin{tabular}{lcccc}
\toprule
$C_{\!\sigma}$ & 10 & 20 & 40 & 80 \\
\midrule
    $E_k$              & 0.0026 & 0.0027 & 0.0028 & 0.0028 \\
    $\varepsilon_s$    & 0.0319 & 0.0166 & 0.0131 & 0.0127 \\
    $\varepsilon_d$    & 0.1318 & 0.0980 & 0.1278 & 0.1433 \\
\bottomrule
\end{tabular}

\end{table}

\paragraph{Grid convergence.}

$E_k$ and $\varepsilon_d$ fall monotonically with resolution.

\begin{table}[htbp]
    \centering
    \caption{Grid convergence at $C_{\!\sigma}=20$, so these entries are a different configuration from the $C_{\!\sigma}=40$ rows of Table~\ref{tab:ctgv_l2}.
    Relative $L_2$ against the $2048^3$ reference over $t_c \in [0,20]$.}
    \label{tab:ctgv_conv}
    \begin{tabular}{lccccc}
\toprule
$N$ & $f_\mu$ & $J_{\min}$ & $E_k$ & $\varepsilon_s$ & $\varepsilon_d$ \\
\midrule
    192 & $3\cdot10^{-2}$ & $3\cdot10^{-3}$ & 0.0091 & 0.0947 & 0.3574 \\   %
    256 & $1\cdot10^{-2}$ & $3\cdot10^{-3}$ & 0.0056 & 0.0412 & 0.3031 \\   %
    320 & $8\cdot10^{-3}$ & $3\cdot10^{-3}$ & 0.0045 & 0.0190 & 0.2746 \\   %
    384 & $1\cdot10^{-2}$ & $3\cdot10^{-3}$ & 0.0037 & 0.0126 & 0.2032 \\   %
    512 & $1\cdot10^{-2}$ & $1\cdot10^{-2}$ & 0.0027 & 0.0166 & 0.0980 \\   %
\bottomrule
\end{tabular}

\end{table}

\paragraph{Dilatational dissipation at $512^3$.}
At $512^3$ the present scheme at $C_{\!\sigma}=40$, the configuration of Table~\ref{tab:ctgv_l2}, ranks first on $\varepsilon_d$ against the seven solvers scored at that same resolution, by a margin of thirty percent over the next best.
It sits inside their spread on $E_k$ and on $\varepsilon_s$ (Table~\ref{tab:ctgv_l2_512}).

\begin{table}[htbp]
    \centering
    \caption{Relative $L_2$ error against the $2048^3$ reference over $t_c \in [0,20]$ at $N=512$.
    The present row is the $C_{\!\sigma}=40$ configuration.}
    \label{tab:ctgv_l2_512}
    \begin{tabular}{lccc}
\toprule
& $E_k$ & $\varepsilon_s$ & $\varepsilon_d$ \\
\midrule
    \textbf{Present} & 0.0028 & 0.0131 & 0.1278 \\
    CODA & 0.0015 & 0.0070 & 0.4289 \\
    NS3D & 0.0009 & 0.0050 & 0.1874 \\
    FLEXI & 0.0005 & 0.0017 & 0.3282 \\
    OpenSBLI & 0.0006 & 0.0045 & 0.4013 \\
    SD3D & 0.0041 & 0.0198 & 0.6826 \\
    SPADE & 0.0040 & 0.0283 & 0.5556 \\
    H3AMR & 0.0059 & 0.0791 & 0.5358 \\
\bottomrule
\end{tabular}

\end{table}

The dilatational dissipation separates the solvers, spanning a factor of three and a half across them, and the sensor is built to control it.
The dilatational dissipation has two peaks, near $t_c = 2.5$ and $t_c = 6.5$, and the scheme reproduces both.
The early peak grows from $55\%$ of the reference amplitude at $256^3$ to $75\%$ at $384^3$ and $88\%$ at $512^3$.
The spectra of Figure~\ref{fig:ctgv_spectrum} stay broadband through the decay.
In the reference the early peak is the taller of the two by a ratio of $1.10$, and the present result approaches that ordering with resolution, from $0.81$ at $256^3$ to $0.98$ at $512^3$.
At $512^3$ with $C_{\!\sigma}=20$ the ratio is $1.01$ and the ordering is recovered.

\paragraph{Subsonic verification.}

Figure~\ref{fig:ctgv_re500} repeats the appendix verification case of Chapelier et al.\ at $\text{Re} = 500$, $M_0 = 0.5$, integrated to $t_c = 10$ as there, where the seven reference solvers collapse onto one another.
They are not all at the same resolution, with three at $256^3$, one at $320^3$, two at $384^3$ and one at $512^3$.
The comparison is therefore against the band they span rather than against a single reference curve.
The shock-free regime admits $f_\mu = 0$, and this case is run there.
At $256^3$ the relative $L_2$ errors against the midline of that band are $0.0018$ in $E_k$, $0.0047$ in $\varepsilon_s$ and $0.0432$ in $\varepsilon_d$.

\section{Conclusion}
A lattice Boltzmann scheme derived automatically from the compressible Navier--Stokes--Fourier equations has been validated against exact solutions and published reference data, beyond its initial MMS-only validation~\cite{pde2lbm}.
The declared target carries the viscous stress and the heat flux as transported state, and the shock sensor is fully local.
On the Sod problem and on Becker's traveling wave the captured shock thickness converges at first order against the exact solutions.
On the supersonic Taylor--Green vortex at $M_0 = 1.25$ the scheme at $512^3$ matches the reference dilatational dissipation more closely than the solvers compared by Chapelier et al.~\cite{Chapelier2024}, by thirty percent over the next best.
It does so with one set of populations per field on the seven-velocity D3Q7 lattice, in single precision.
A scheme compiled from the declared conservation laws, reading one cell per update, thereby stands alongside hand-built high-order solvers on a supersonic benchmark, and ahead of them on the measure their spread is widest on.

As all cases here are periodic an obvious next step is to generate from the same declaration the boundary treatment that physical geometries require, extending the validation given here from periodic boxes to wall-bounded compressible flow.

\section*{Code Availability}

The declared system is given in full by Listing~\ref{lst:declaration}, the shock capturing by Section~\ref{sec:sensor}, and the parameters of both by Table~\ref{tab:parameters}, so every result reported here is reproducible from the stated values.
The \emph{PDE2LBM} compiler is available upon reasonable request.

\section*{Acknowledgement}

Google Gemini and Anthropic Claude assisted in drafting and revising the manuscript.
All content was reviewed, verified, and approved by the authors, who take full responsibility for it.

\bibliographystyle{unsrturlabbrv}
\bibliography{bibliography_cnse}

\end{document}